\documentclass[pra,
reprint,
footinbib,
notitlepage,
twocolumn
]{revtex4-2}

 \usepackage{graphicx}
 \usepackage{dcolumn}
 \usepackage{bm}
 \usepackage{bbm}
 \usepackage{mathrsfs}
 \usepackage{hyperref}
 \usepackage[mathlines]{lineno}
 
 \usepackage{url}
 \usepackage[dvipsnames]{xcolor}
 \usepackage{amsthm}
 \usepackage{slashed}
 \usepackage{amsmath, mathtools}
 \usepackage{amsfonts}
 \usepackage{graphicx}
 \usepackage{hyperref}
 \usepackage{comment}
 
 \hypersetup{
 	colorlinks   = true, 
 	urlcolor     = blue, 
 	linkcolor    = blue, 
 	citecolor   = red 
 }

 \newcommand{\eqnref}[1]{Eq.~(\ref{#1})}
 \newcommand{\figref}[1]{Fig.~\ref{#1}}
 \newcommand{\sfigref}[2]{Fig.~\hyperref[#1]{\ref{#1}#2}}

 \newcommand{\abs}[1]{\left| #1 \right|}

 \newcommand{\beq}{\begin{equation}}
 	\newcommand{\eeq}{\end{equation}}
 \newcommand{\beqd}{\begin{equation*}}
 	\newcommand{\eeqd}{\end{equation*}}
 \newcommand{\dou}{\partial}
 \newcommand{\up}{\uparrow}
 \newcommand{\down}{\downarrow}

 \newcommand{\bpm}{\begin{pmatrix}}
 	\newcommand{\epm}{\end{pmatrix}}
 \newcommand{\sx}{\sigma_{x}}
 \newcommand{\sy}{\sigma_{y}}
 \newcommand{\sz}{\sigma_{z}}

 \newcommand{\id}[1]{\mathbbm{1}_{#1}}

 \newcommand{\bcen}{\begin{center}}
 	\newcommand{\ecen}{\end{center}}
 \newcommand{\btab}{\begin{tabular}}
 	\newcommand{\etab}{\end{tabular}}
 \newcommand{\bdes}{\begin{description}}
 	\newcommand{\edes}{\end{description}}

 \newcommand{\bea}{\begin{eqnarray}}
 	\newcommand{\eea}{\end{eqnarray}}

 \newcommand{\bary}{\begin{array}}
 	\newcommand{\eary}{\end{array}}
 \newcommand{\benum}{\begin{enumerate}}
 	\newcommand{\eenum}{\end{enumerate}}
 \newcommand{\bitem}{\begin{itemize}}
 	\newcommand{\eitem}{\end{itemize}}

 \newcommand{\bsig}{{\boldsymbol{\sigma}}}

 \newcommand{\ba} {{\boldsymbol{a}}}

 \newcommand{\bd} { {\boldsymbol{d}} }

 \newcommand{\bk} { {\boldsymbol{k}} }

 \newcommand{\bq} {{ \boldsymbol{q}} }

 \newcommand{\bu} { {\boldsymbol{u}} }

 \newcommand{\bz} { \mbox{\boldmath $z$}}

 \newcommand{\bK} { \boldsymbol{K} }

 \newcommand{\bQ} { \boldsymbol{Q} }

 \newcommand{\cH}{\mathscr{H}}

\newcommand{\SMSec}[1]{S#1}

 \newcommand{\Integers}{{\mathbb{Z}}}
  \newcommand{\ci}{\mathbbm{i}}
\newcommand{\mytitle}{\protect{{Hopf Semimetals}}}

 \newcommand{\SMtopoindex}{1}
 \newcommand{\SMhopfins}{2}
 \newcommand{\SMhopfsemimetal}{3}
 \newcommand{\SMsurfacestate}{4}

\begin{document}

	\title{\mytitle}

	
	\author{Bhandaru Phani Parasar}
	\email{bhandarup@iisc.ac.in}
	\affiliation{Centre for Condensed Matter Theory, Department of Physics, Indian Institute of Science, Bangalore 560012, India}
	\author{Vijay B.~Shenoy}
	\email{shenoy@iisc.ac.in}
	\affiliation{Centre for Condensed Matter Theory, Department of Physics, Indian Institute of Science, Bangalore 560012, India}

\begin{abstract}
    We construct two-band topological semimetals in four dimensions using the unstable homotopy of maps from the three-torus $T^3$ (Brillouin zone of a 3D crystal) to the two-sphere $S^2$. Dubbed ``Hopf semimetals'', these gapless phases generically host nodal lines, with a surface enclosing such a nodal line in the four-dimensional Brillouin zone carrying a Hopf flux. These semimetals show a unique class of surface states: while some three-dimensional surfaces host gapless Fermi-arc states {\em and} drumhead states, other surfaces have gapless  Fermi surfaces. Gapless two-dimensional corner states are also present at the intersection of three-dimensional surfaces. 
\end{abstract}

\maketitle

\noindent
\underline{\em Introduction:} The understanding and classification of gapped topological phases\cite{KaneTI3D,KaneQSHE,MooreBalents2007,Roy2009,KitaevPT,RyuTFW, QiZhang2011,ChiuRyu2016} of non-interacting fermions has not only provided deeper insights, 
but also, stimulated wider generalizations\cite{Slager2013,Benalcazar2019} and the search for topological materials\cite{Bradlyn2017}. 
The current understanding of these gapped phases is built on 
the symmetry classification of the fermionic systems 
that arise from the presence or absence of intrinsic symmetries such as time reversal, charge conjugation and sublattice symmetries\cite{Altland1997,Zirnbauer2010,ChiuRyu2016,Agarwala2017}. 
In a crystalline system in $d$-dimensions, the ground state of a gapped fermionic system is obtained by the state of occupied valance bands 
in the first Brillouin zone (BZ), the $d$-torus $T^d$. Interestingly, the occupied states at any point in the BZ can be viewed as a point in one of the ten symmetric spaces ${\cal S}$, the specific one being determined by the intrinsic symmetry.  Topologically distinct gapped ground states are identified with the homotopy classes of maps from $T^d$ to ${\cal S}$,  resulting in the periodic table of strong topological phases\cite{KitaevPT}.

Apart from these symmetry protected topological phases, a class of gapless phases have elicited attention, beginning with graphene\cite{CastroNeto2009}, and more recently, Weyl and Dirac semimetals \cite{Burkov2011,Hosur2013,Vafek2014,Fu2015,BinghaiFelser2017,Armitage2018,Burkov2018}. Weyl semimetals arise in three dimensions, exploiting the topology in a lower dimensional slice of the Brillouin zone (say the $k_1-k_2$ plane) that undergoes a ``phase transition'' as the $k_3$ of the slice is varied. Thus, these semimetals are protected by the topology of the two adjacent two-dimensional phases, the gapless points being those $k_3$ at which the quantum phase transition between the two-dimensional phases is affected.  They have received considerable attention owing to the exotic properties such as Fermi-arc surface states, interesting nonlinear responses related to the chiral anomaly etc. 

The classification of gapped phases hinges on the number of bands being large. In more mathematical terms, these are determined by the stable homotopies of maps from $T^d$ to ${\cal S}$ which are realized when $\cal S$  is large dimensional. In the absence of a large number of bands, one can still obtain topological phases that arise from unstable homotopies of maps from $T^d$ to ${\cal S}$ i.~e., when the space ${\cal S}$ is ``small dimensional''. An example in a three-dimensional lattice that hosts a two-band gapped system is dubbed as a ``Hopf insulator''\cite{Moore2008,DengDuan2013,Kennedy2016} whose topology can be traced to the homotopies of maps from the three-sphere $S^3$ to the two sphere $S^2$. 

In this paper, we show how a Hopf insulator in 3 dimensions can be used to construct interesting gapless phases in  4 dimensions.  These gapless phases have several new features. Unlike the Weyl semimetal in three dimensions, these four-dimensional semimetals host nodal lines of gapless points (a one-dimensional submanifold) in the four-dimensional Brillouin zone. Remarkably, any three-dimensional surface that encloses one of these rings carries an integer Hopf number that characterizes the phases on either side of these rings. These features manifest spectacularly in the nature of gapless surface states. There are three-dimensional surfaces, which host  Fermi-arc states and, in addition, gapless drumhead states \cite{Hasan2016}. Further, we also find evidence of two-dimensional corner states that arise at the intersection of two three-dimensional surfaces of the four-dimensional insulator.  This work presents a new class of interesting topological phases in higher dimensions\cite{Price2015,Lian2016,Petrides2018}.

\noindent
\underline{\em Hopf and Hopf-Chern Insulators:} We begin with a two-band system that realizes an insulting phase on a 3D cubic lattice with a unit lattice spacing. The Brillouin zone (BZ) of this system is the three torus $T^3$ corresponding to $[-\pi,\pi]^3$. A generic point in the BZ is denoted by $\bk = (k_1,k_2,k_3)$. A  two-band Hamiltonian is defined by 
\beq \label{eqn:twobandH}
H(\bk) = \bd(\bk) \cdot \bsig
\eeq
where $\bd(\bk)$ is the vector $(d_1(\bk),d_2(\bk),d_3(\bk)$), and $\bsig = (\sigma_1,\sigma_2,\sigma_3)$ where $\sigma_i$ are the $2 \times 2$ Pauli matrices. The chemical potential here and henceforth in this paper is set to zero so that the fermionic many-body system is half-filled ($1$ particle per site). Existence of a gap necessitates that $|\bd(\bk)| > 0$ for $\bk \in T^3$, and thus the unit vector $\hat{d}(\bk) = \bd(\bk)/|\bd(\bk)|$ can be identified with a point on the two-sphere $S^2$. Consequently, the Hamiltonian \eqnref{eqn:twobandH} can be viewed as a map from $T^3$ to $S^2$.

Distinct insulating topological phases are obtained depending on the homotopy class of the map from $T^3$ to $S^2$, with two insulators being identical if they can be smoothly deformed to each other (i.e., homotopic) without closing the gap. Such maps have been extensively studied both from the mathematical and physical perspectives\cite{Pontrjagin1941,Auckly2005,DeTurck2013,Kobayashi2013,Unal2019}. The homotopy classes of the maps are characterized by four (integer) numbers $(\chi,(C_1,C_2,C_3))$. The numbers $C_\alpha$ are the Chern numbers associated with two-dimensional $T^2$ submanifolds of $T^3$, where $\alpha$ indicates the normal direction to the $T^2$-submanifold. 
The number $\chi$ is in $\Integers_{2 Q}$ where $Q = \textup{GCD}(C_1,C_2,C_3)$. Thus, if all $C_\alpha$ are zero, $\chi$ can take any integer value, and such insulators are termed as Hopf insulators\cite{Moore2008,DengDuan2013,Kennedy2016}. On the other hand, if any of the $C_\alpha$ is nonzero, then $\chi$ takes on only a finite set of values, and such insulators are dubbed as Hopf-Chern insulators\cite{Kennedy2016}. 

Hopf insulators can be constructed\cite{Moore2008,DengDuan2013} using an intermediate map from $T^3$ to $S^3$ (the three-sphere). Since $S^3$ is described by two complex number $z_1,z_2$ such that $|z_1|^2 + |z_2|^2 > 0$, the prescription
\beq\label{eqn:T3toS3}
\begin{split}
z_1(\bk,h) &= \sin k_1 + \ci  \sin k_2 \\
z_2(\bk,h) &= \sin k_3 + \ci (\cos k_1 + \cos k_2 + \cos k_3 + h)
\end{split}
\eeq
($h$ is a parameter, $\ci = \sqrt{-1}$), is a map from $T^3$ to $S^3$. The topological index  $\Gamma$ (\cite{SM}, section \SMSec{\SMtopoindex}) of this map, vanishes when $|h|>3$, is $1$ for $1 < |h| < 3$, and $-2$ for $|h| < 1$. The map is thus topologically nontrivial for $|h| < 3$. Finally, to obtain a two-band model, the point on $S^3$ is mapped to $S^2$ via the Hopf map\cite{BottTu1982}
\beq\label{eqn:HopfS3toS2}
\begin{split}
\bd^{(p,q)}(\bk,h) = & \left(2 \Re{(z_1^p(\bk,h) z_2^{*q}(\bk,h))},  \right. \\  
 2 \Im{(z_1^p(\bk,h) z_2^{*q}(\bk,h))}, & \left. |z_1(\bk,h)|^{2p} - |z_2(\bk,h)|^{2q} \right)
\end{split}
\eeq
where $p,q$ are co-prime integers, $*$ denotes complex conjugation. Such a map has a Hopf index\cite{BottTu1982} $\cH = \pm pq$. The Hopf insulator defined using \eqnref{eqn:HopfS3toS2} has vanishing Chern number $C_\alpha$, and is thus characterized by $(\chi,(0,0,0))$ where $\chi= \Gamma \cH$\cite{DengDuan2013}. 

Hopf-Chern insulators are those which have a non-zero Chern numbers $C_\alpha$. These are obtained\cite{Kennedy2016,WangPallab2023} using
\beq\label{eqn:3DHopfChern_hamiltonian}
\begin{split}
\begin{pmatrix}
    d^{(m)}_1(\bk) \\
    d^{(m)}_2(\bk)
\end{pmatrix}
&=
\begin{pmatrix}
    \cos{m k_1} & - \sin{m k_1} \\
    \sin{m k_1} & \cos{m k_1}
\end{pmatrix}
\begin{pmatrix}
    \sin{k_2} \\
    \sin{k_3}
\end{pmatrix}\\
d^{(m)}_3(\bk)=1 + &\Delta_1 (\cos{k_2}+\cos{k_3}) +\Delta_2 \cos{k_2} \cos{k_3}
\end{split}
\eeq
where $m$ is an integer and $\Delta_1$,$\Delta_2$ are real parameters. For this model, $C_2=C_3=0$ always, and $C_1$ is determined by the values of $\Delta_1$ and $\Delta_2$ (and is not affected by the value of $m$). The quantity $\chi$ is determined by $m$ as $\chi = m |C_1| \textup{mod} \,2 |C_1|$.

\begin{figure}
    \centering
    \includegraphics[width=\linewidth]{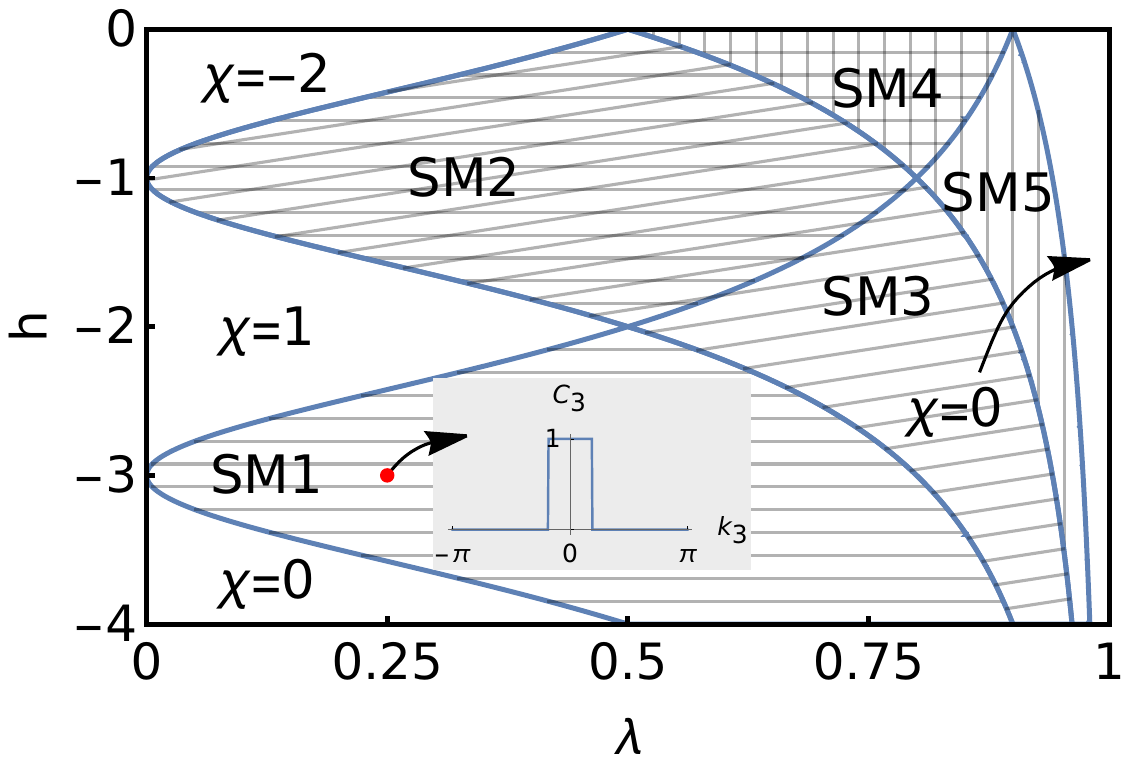}
    \caption{Phase diagram of  system in \eqnref{eqn:Hopfhl}. Semimetallic regions with horizontal lines have band touchings at $(0,0,\pm k_t^{(1)})$, those with vertical lines have touchings at $(\pi,\pi,\pm k_t^{(-1)}$),  and those with slanted lines have touching at $(\pi,0,\pm k_t^{(0)}),(0,\pi,\pm k_t^{(0)})$, where $k^{l}_t = \arccos{\left( \frac{\frac{\lambda}{1-\lambda} -1 - (h + 2 l)^2}{2 (h + 2l)}\right)}$. Inset: Chern number of $T^2$ submanifold of $T^3$ BZ labelled  by $k_3$, for $h=-3, \lambda= 1/4$. }
    \label{fig:HopfPD}
\end{figure}

\medskip
\noindent
\underline{\em Hopf semimetals:} 
Realization of such topological phases in $d$ dimensions allows us to construct interesting semimetallic phases in $d+1$ dimensions. In fact, the well-studied Weyl semimetallic phases are examples of such physics. These phases in three dimensions arise from the topological Chern insulators in two dimensions and enjoy a degree of protection owing to the stability of Weyl points where $d_i(\bk) = 0$. The Weyl points are those at which all components of $\bd$ vanish, each such equation describing a surface embedded in three dimensions. Three such surfaces generically intersect at isolated points, leading to the stability of Weyl points to small perturbations of the Hamiltonian. Taking this idea to a four-dimensional two-band system, semimetallicity will require that $d_i(\bK) = 0$ for $i=1,2,3$ where the gap closes; $\bK$ is a point in the BZ $T^4$ (four-torous, $[-\pi,\pi]^4$) of the 4D cubic lattice with $\bK = (k_1,k_2,k_3,k_4) \equiv (\bk,k_4)$. This condition, if satisfied, will be generically met on a one-dimensional submanifold of $T^4$. The conclusion is that the semimetals arising in these four-dimensional systems will generically possess {\em nodal lines}. The exciting aspect here is that these line nodes enjoy a degree of protection in that small perturbations cannot remove them but, at best, change their shape. 

To construct such a semimetal arising from the topology of the Hopf insulator, we first study a quantum phase transition that occurs in the 3D Hopf insulator. Consider a three-dimensional  system with a tuning parameter $\lambda$
\beq\label{eqn:Hopfhl}
\bd(\bk,h,\lambda) = (1-\lambda)\bd^{(1,1)}(\bk,h) + \lambda \bd^{\textup{f}}(\bk)
\eeq
where $\bd^{(1,1)}(\bk)$ is the dispersion in \eqnref{eqn:HopfS3toS2}, and $\bd^{\textup{f}}(\bk) = (0,0,1)$ is the dispersion of a gapped flat band system which is topologically trivial. When $\lambda = 0$, the system hosts Hopf insulating phases in a regime of the parameter $h$, and a topologically trivial phase for $\lambda=1$. For intermediate values of $\lambda$  we obtain a variety of semimetallic phases (see \figref{fig:HopfPD}).  These semimetallic phases are characterized by a change in Chern numbers of the $T^2$ submanifolds of $T^3$ (\cite{SM}, section \SMSec{\SMhopfins}), as illustrated for $\lambda = 1/4$ in \figref{fig:HopfPD} (inset). 


We construct a semimetallic phase on a four-dimensional cubic lattice by defining for each $\bK \in T^4$ 
\beq\label{eqn:HopfNLSM}
\bd(\bK,\lambda) = (1-\lambda) \bd^{(1,1)}(\bk,-3 + \cos{k_4}) + \lambda \bd^{\textup{f}}(\bk) 
\eeq
The Hamiltonian \eqnref{eqn:twobandH} obtained using this produces a semimetallic phase for a range $0 \le \lambda < 1/2$. Focusing first on $\lambda=0$, we find that the bulk gap closes at {\em two points} in $T^4$, namely $ \bK_H^\pm (0,0,0,\pm \pi/2)$, where the bands  touch quadratically (\cite{SM}, section \SMSec{\SMhopfsemimetal}).  Most interestingly, these points are a source of ``Hopf flux'' in $T^4$; this is most easily seen by enclosing, for example, the point $\bK^+$ by a ball $B^+ = |\bK - \bK^+| \le \epsilon$ where $\epsilon$ is a small number, the boundary of this ball $\dou B^+$ is homeomorphic to $S^3$, and $\bd(\bK,0), \bK \in \dou B^+ $ defines a map from $S^3$ to $S^2$. Interestingly, the map carries a non-vanishing Hopf index $\cH = -1$! (\cite{SM}, section \SMSec{\SMhopfsemimetal}), pointing to the topological nature of this ``Hopf semimetal'' similar to what is found in a three-dimensional Weyl semimetal. 

\begin{figure}
    \centering
    \includegraphics[width=\linewidth]{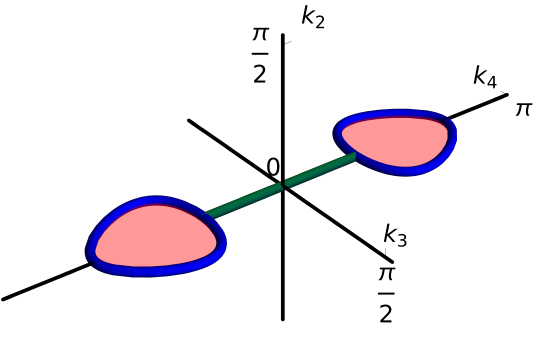}
    \caption{Hopf nodal line semimetal for $\lambda = 1/4$ in \eqnref{eqn:HopfNLSM}. Blue: nodal lines (nodal lines lie in the $k_1=0$ submanifold of $T^4$). The figure also depicts $T^3$ the {\em surface BZ} of the $(1,0,0,0)$ surface. Green: Fermi arc surface states that correspond to the $(1,0,0)$ surface states of the $\chi=1$ Hopf insulator. Red: Drumhead surface states that correspond to the edge states of the $C_3=1$ Chern insulator. }
    \label{fig:HopfNLandSBZ1000}
\end{figure}

However, as noted in the discussion above, point touching of two bands in 4 dimensions is not stable (in contrast to the 3D Weyl semimetal), and this is indeed seen in our construction. For a small $\lambda > 0$, we find that the two Hopf points evolve to nodal lines where the bands touch linearly except at two points on the nodal line (\cite{SM}, section \SMSec{\SMhopfsemimetal}). With increasing $\lambda$, the size of the nodal lines centered around $\bK_\pm$ increases. \figref{fig:HopfNLandSBZ1000} (thick blue lines) shows the nodal lines for $\lambda=1/4$ in the $k_1=0$ $T^3$ sub-manifold of the $T^4$ BZ. The nodal lines appear in the $k_3-k_4$ plane ($k_1=k_2=0$), and encircle the Hopf points $\bK^\pm_H$ extending from $k_4^{\textup{min}} \leq |k_4| \leq k_4^{\textup{max}}$,
$k_4^{\textup{min}}=\arccos{\sqrt{\frac{\lambda}{1-\lambda}}},
k_4^{\textup{max}}=\pi-\arccos{\sqrt{\frac{\lambda}{1-\lambda}}}$
and described by the equation 
\beq\label{eqn:EqnOfNodalLine}
2(1-\cos{k_3})(1-\cos{k_4})+\cos^2{k_4}=\frac{\lambda}{1-\lambda}.
\eeq
The nodal lines $L^{\pm}$ respectively encircle $\bK_H^{\pm}$.
The intriguing aspect is that the nodal lines also carry the same Hopf number, i.e.~if we place  balls $B^\pm$ centered around $\bK^\pm_H$, and {\em enclosing} the nodal lines $L^{\pm}$, then the Hamiltonian on the surface of the ball $\dou B^\pm$ defines a Hopf map such that the Hopf invariant associated with the $L^\pm$ nodal lines are opposite of each other. This demonstrates the topological origin of the nodal lines and their stability. The nodal lines which appear between $\pm k_4^{\textup{min}}$ and $\pm k_4^{\text{max}}$ separate three dimensional $T^3$ sub-manifolds of $T^4$ that carry distinct invariants $\chi$. Indeed, for all the $T^3$ submanifolds with $|k_4| < k_4^{\textup{min}}$, the invariant $\chi=1$. This change of topology of the bands along $k_4$ is encoded in the Hopf number on the surface of $\dou B^{\pm}$.

\begin{figure}
    \centering
    \includegraphics[width=0.7\linewidth]{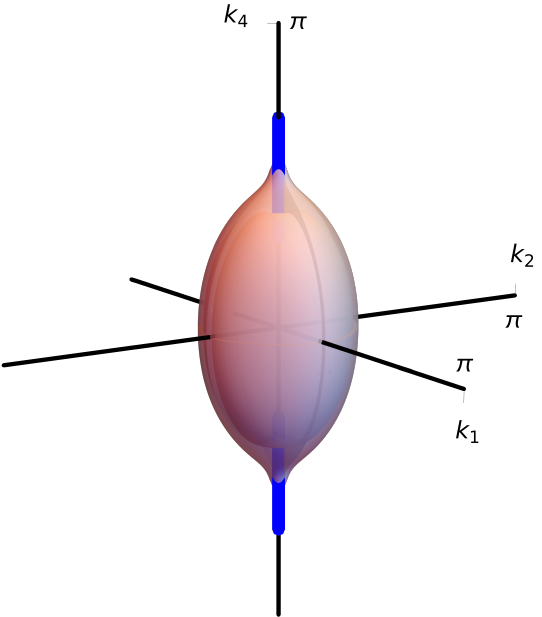}~
    \caption{Surface states of the Hopf nodal line semimetal in the $(0,0,1,0)$ surface BZ. Red: ``Fermi surface'' states that correspond to the $(0,0,1)$ surface states of the $\chi=1$ Hopf insulator. Blue: Fermi arc states that arise from the projection of the nodal lines onto the surface BZ. }
    \label{fig:Hopf_surface_dir3}
\end{figure}

We next investigate the nature of the surface states of the four-dimensional Hopf semimetal. The surface of this system is characterized by a normal direction, and is a ``three-dimensional crystal'' with a $T^3$ surface Brillouin zone. Depicted in \figref{fig:HopfNLandSBZ1000} for the surface with the $(1,0,0,0)$ normal, are a remarkably rich set of surface states. First, there is a set of gapless ``Fermi-arc'' states that exist between $\pm k_4^{\textup{min}}$, depicted by the solid green line in \figref{fig:HopfNLandSBZ1000}. These arise from the $(1,0,0)$ surface states of the $\chi=1$ Hopf insulator realized in the $T^3$ submanifolds in this regime of $k_4$. There are additional surface states that arise in the regime $k^{\textup{min}}_4 < |k_4| < k^{\textup{max}}_4$. In fact, all the points in the $T^3$ surface BZ that are inside the nodal line projected onto the surface BZ host gapless states that are higher-dimensional analogs of drumhead states(\cite{Hasan2016} and references therein). Details of all of these states may be found in \cite{SM}, section \SMSec{\SMsurfacestate}.

The Hopf semimetal holds further interesting aspects when we study the surface states on the $(0,0,1,0)$ surface. This surface hosts two types of gapless states (see \figref{fig:Hopf_surface_dir3}). We find, first, a ``Fermi surface'' of gapless states between $|k_4| < k_4^{\textup{min}}$; these are the surface states of the $\chi=1$ Hopf insulator that is realized in the $T^3$ submanifolds. In addition, there are other gapless states shown by the blue lines of the same figure; these are gapless states corresponding to the projection of the gapless nodal line on to the surface BZ. 

Finally, we also point out the possibility of interesting ``corner states'' in this Hopf semimetal that arise in the two-dimensional intersection of two three-dimensional surfaces. As an example, The corner formed by the intersection of two surfaces (1,0,0,0) and (0,1,0,0), will have a two-dimensional $T^2$ Brillouin zone labeled by $(k_3,k_4)$. The corner states arise because the corner terminates $1-2$ planes of the four-dimensional crystal. In the instance of $\lambda=1/4$, some of the $1-2$ submanifolds (which are $T^2$) host non-zero Chern numbers in the regime $k_4^{\textup{min}}<|k_4| < k_4^{\textup{max}}$, and should result in the ``corner drumhead states'' in the $T^2$ Brillouin zone. Other corners (intersections of different three-dimensional surfaces) will host Fermi arc states. While our calculations are consistent with this possibility, a full demonstration of this requires very large system sizes.

\begin{figure}
    \centerline{\includegraphics[width=0.5\linewidth]{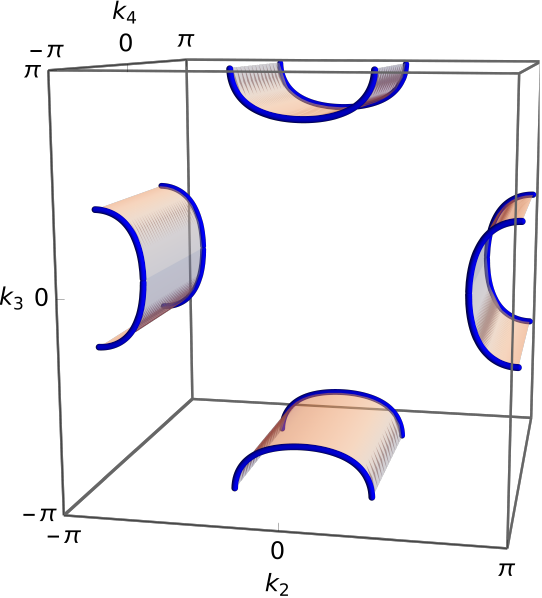}~\includegraphics[width=0.5\linewidth]{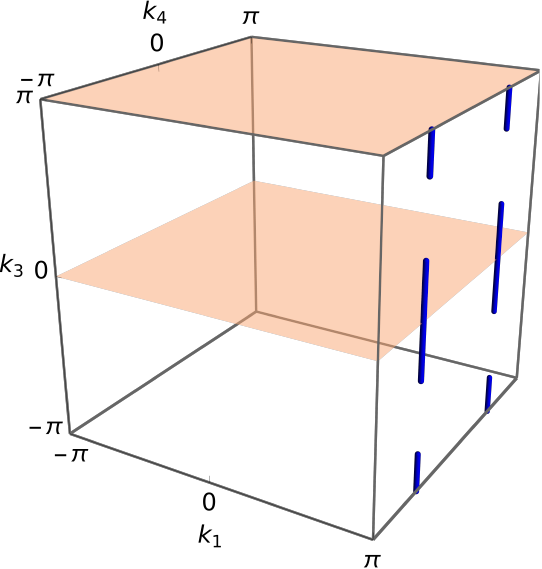}}
    \centerline{{\bf (a)}~~~~~~~~~~~~~~~~~~~~~~~~~~~~~~~~~{\bf (b)}}
    \caption{ {\bf (a)} Hopf-Chern nodal line semimetal. Blue: Nodal lines that lie in the $T^3$ submanifold with  $k_1=\pi$  of $T^4$. The figure also depicts the $(1,0,0,0)$ surface BZ. Red: ```Fermi-cylinder'' states that correspond to surface states of $(1,0,0)$ surface of the $(2 mod 4, (2,0,0))$ Hopf-Chern insulator. {\bf (b)} $(0,1,0,0)$ surface states of the Hopf-Chern nodal line semimetal. They correspond to the edge states of $C=2$ Chern insulator. Here, the blue lines depict the projection of the nodal lines on to the surface Brillouin zone.  }
    \label{fig:HopfC_surface_dir1}
\end{figure}

In the last part of this paper, we demonstrate that such semimetals in four dimensions can also be constructed out of the Hopf-Chern insulators. To achieve this we first construct as 3d system with a parameter $\lambda$,
\beq
\bd(\bk,\lambda) = (1 - \lambda) \bd^{(1)}(\bk) + \lambda  \bd^{(0)}(\bk)
\eeq
where $\bd^{(1)}$ and $\bd^{(0)}$ are obtained from \eqnref{eqn:3DHopfChern_hamiltonian} with $3 \Delta_1 = \Delta_2 = \frac{3}{2}$. This system undergoes a quantum phase transition from a Hopf-Chern insulator with $(\chi = 2, (C_1=2,C_2=0,C_3=0))$ to another insulator with $(\chi=0, (C_1=2,C_2=0,C_3=0)$ which can be viewed as a stack of Chern insulators. The phase transition\footnote{We note that a more generic version of this problem will have a gapless regime of $\lambda$ separating the two phases rather than a single quantum critical point.} occurs at $\lambda=1/2$ where the gap closes along two nodal-line rings, and $\chi$ changes from $2$ to $0$ for $\lambda > 1/2$, with no change in the Chern numbers, i.~e., $C_1=2$ for $\lambda \ne 1/2$ ($C_{2,3}= 0$). 

We use the above to construct a Hopf-Chern semimetal in a four dimensional cubic lattice, via
\beq\label{eqn:HopfChern}
\bd(\bK) = \left( \frac{1- \cos{k_4}}{2} \right) \bd^{(1)}(\bk) +  \left( \frac{1 + \cos{k_4}}{2} \right)\bd^{(0)}(\bk)
\eeq
where $\bK=(\bk,k_4)$ is a point in BZ $T^4$. This system hosts nodal lines in the $k_2-k_3$ planes with $k_4 = \pm \frac{\pi}{2}$ as shown by  blue lines in \figref{fig:HopfC_surface_dir1}(a). The $T^3$ submanifolds of $T^4$ with $|k_4| < \pi/2$ have Hopf-Chern character with $\chi=2$.  The topological aspects of this semimetal are again evident in the nature of the surface states that it hosts. On a surface with $(1,0,0,0)$ as the normal, we find a ``Fermi-cylinder'' of gapless states (see \figref{fig:HopfC_surface_dir1}(a)). The end-arcs of these cylinders are the projection of the nodal lines onto the surface BZ. These arise from the surface states of the Hopf-Chern insulators residing in  $|k_4| < \pi/2$ region of  $T^4$ BZ. Turning now to the $(0,1,0,0)$ surface (see \figref{fig:HopfC_surface_dir1}(b)), we find that there are two types of gapless states. Since this also terminates at $2$-direction of the crystal, it terminates $2-3$ planes of the crystal, which carry a Chern number (independent of the value of $k_4$). Thus, there is a set of gapless states on the planes with $k_3=0$ and $k_3=\pi$. Finally there is a second set of gapless states that arise as disjoint nodal lines on the $k_1=\pi$ plane of the surface $BZ$; these states are the projections of the nodal lines on to the surface BZ. It is also clear that this system can host a variety of corner states, states residing on the two-dimensional intersection of two three-dimensional surfaces.


 It is interesting to explore the possibilities of experimental realization of these four-dimensal Hopf-semimetals and their surface states exploiting the ideas of synthetic dimensions\cite{Ozawa2019} in cold atoms\cite{Lohse2018} and photonic systems\cite{Zilberberg2018}. Further theoretical investigations should also be fruitful. It will be interesting to find generalizations of such semimetals using other recently proposed topological phases\cite{Davoyan2023} in three dimensions. Understanding the responses\cite{Sekine2021, WangPallab2023} of the Hopf semimetals also provides an exciting direction.

\noindent
{\em Acknowledgement:} BPP thanks the PMRF program for support. VBS acknowledges DST-SERB, India, for support through a MATRICS grant.

\bibliographystyle{apsrev4-2}
\bibliography{ref}

\def\makeSM{0}
\ifdefined\makeSM

\newwrite\tempfile
\immediate\openout\tempfile=junkSM.\jobname
\immediate\write\tempfile{\thepage }
\immediate\closeout\tempfile

\clearpage
\newpage
\appendix

\renewcommand{\appendixname}{}
\renewcommand{\thesection}{{S\arabic{section}}}
\renewcommand{\theequation}{\thesection.\arabic{equation}}
\renewcommand{\thefigure}{S.\arabic{figure}}
 
\setcounter{page}{1}
\setcounter{figure}{0}

\begin{widetext}

\maketitle

\centerline{\bf Supplemental Material}
\centerline{\bf for}
\centerline{\bf \mytitle}
\medskip
\centerline{by Bhandaru Phani Parasar and Vijay B.~Shenoy}
\bigskip
\end{widetext}

\section{Topological Index of Map from $T^3$ to $S^3$}
\label{sec:supp_GammaT3toS3}
A gapped two-band hamiltonian in three dimensions can be
	 thought of a map from the three-torus $T^3$ to the two-sphere $S^2$. A tight binding model for Hopf insulator on a cubic lattice in three dimensions can thus be constructed \cite{Moore2008, DengDuan2013} by first mapping a point on $T^3$ to a point on $S^3$ (\eqnref{eqn:T3toS3}, main text). This point is then mapped to a point on $S^2$ via the Hopf map (\eqnref{eqn:HopfS3toS2}, main text). The $T^3$ to $S^3$ map is given by
\beq\label{eqn:supp_T3toS3}
\begin{split}
z_1(\bk,h) &= \sin k_1 + \ci  \sin k_2 \\
z_2(\bk,h) &= \sin k_3 + \ci (\cos k_1 + \cos k_2 + \cos k_3 + h)
\end{split}
\eeq
where $h$ is a real parameter such that $|z_1|^2 + |z_2|^2 > 0 \; \forall \bk \in T^3$. The homotopic classification of maps from $T^3$ to $S^3$ is $\Integers$ and the homotopic invariant $\Gamma$ encodes the winding of $T^3$ on to $S^3$. For the map given above, it is
\beq\label{eqn:supp_Gamma}
\Gamma=\frac{1}{12 \pi^2} \int_{\text{BZ}} d \bk \; \epsilon_{\alpha \beta \gamma \rho} \epsilon_{\mu \nu \tau }\frac{1}{|\bz|^4}\bz_{\alpha} \partial_\mu \bz_{\beta} \partial_\nu \bz_{\gamma} \partial_\tau \bz_{\rho}
\eeq
where $\bz=\left(\Re{\left(z_1\right)},\Im{\left(z_1\right)},\Re{\left(z_2\right)},\Im{\left(z_2\right)}\right)$. The homotopic invariant depends on the parameter $h$ according to
\begin{align}\label{eqn:supp_Gamma_values}
	\Gamma =\left\{
		\begin{array}{cc}
			0 & \abs{h}>3 \\
			1 & 1<\abs{h}<3 \\
			-2 & \abs{h}<1
		\end{array}
	\right.
\end{align}
Finally, the $T^3$ to $S^2$ map is given by composing the above with the Hopf map to give:
\begin{align}\label{eqn:supp_T3toS2}
\nonumber
\bd^{(p,q)}(\bk,h) = &\left(2 \Re{(z_1^p(\bk,h) z_2^{*q}(\bk,h))}, 2 \Im{(z_1^p(\bk,h) z_2^{*q}(\bk,h))}, \right. \\ & \left. |z_1(\bk,h)|^{2p} - |z_2(\bk,h)|^{2q} \right)
\end{align}
Here $p,q$ are co-prime integers.
For a map from $T^3$ to $S^2$ constructed this way, it can be shown \cite{DengDuan2013} that the Chern numbers $C_1$, $C_2$, $C_3$ associated with the $T^2$ submanifolds of $T^3$ vanish. Thus the homotopy of this map is characterized by $\chi$, which is equal to the product of the homotopic invariant from $T^3$ to $S^3$ ($\Gamma$) and the Hopf index from $S^3$ to $S^2$ ($\cH =\pm pq$): $\chi= \Gamma \cH$

\section{Phase transition in 3D Hopf Insulator}\label{sec:supp_hopfins_pd}
To illustrate the nature of phase transition between distinct Hopf insulators, and its implications for Hopf semimetals, we consider the following two-band system in three dimensions, with real parameters $\lambda, h$ (\eqnref{eqn:Hopfhl}, main text) :
\begin{equation}\label{eqn:supp_hopfhl}
\bd(\bk,h,\lambda) = (1-\lambda)\bd^{(1,1)}(\bk,h) + \lambda \bd^{\textup{f}}(\bk)
\end{equation}
Here, $\bd ^{\left(1,1\right)}\left(\bk, h\right)$ is as given in \eqnref{eqn:supp_T3toS2}, and $\bd^{\textup{f}}(\bk)=(0,0,1)$ corresponds to a flat band dispersion. The phase diagram of this system in $\lambda-h$
 plane is given in \figref{fig:HopfPD}, main text. When $\lambda=0$, the system is a trivial insulator for $h<-3$, Hopf insulator with $\chi=1$ for $-3<h<-1$, and a Hopf insulator with $\chi=-2$ for $h>-1$. i.e., if $\lambda=0$ is fixed and $h$ is varied, there is a quantum phase transition occurring at $h=-3$ between $\chi=0$ and $\chi=1$ insulators. The semimetal at the critical point has quadratic band touching at $\bk^0 = \left(0,0,0\right)$. To see this, we expand the Bloch hamiltonian to the lowest non-trivial order in $q=\bk -\bk^0$:
\begin{equation}\label{eqn:supp_lowE__hopfcritical}
    H(\bk^0+\bq) \approx (q_1^2+q_2^2-q_3^2) \sigma_3 +2 q_3 (q_1 \sigma_1 + q_2 \sigma_2)        
\end{equation}
Note that as $h$ is varied, the transition between distinct Hopf insulators occurs at a sharply defined value of $h$ only for the finely tuned value of $\lambda=0$. Generically, the transition between distinct Hopf insulators involves a region of semimetal.

Now, we identify various semimetallic regions shown in \figref{fig:HopfPD}, main text. It is immediately seen that the spectrum of \ref{eqn:supp_hopfhl} has a vanishing gap at $\bk$ if
\begin{align}
    z_1(\bk)&=0\\
    |z_2(\bk)|^2&=\frac{\lambda}{1-\lambda}
\end{align}
The possible solutions are
\begin{align}\label{eqn:supp_kt_solns}
\nonumber
    \bk&=(0,0,\pm k_t^{(1)}); k_t^{(1)}= \arccos\left(\frac{\frac{\lambda}{1-\lambda}-1-(h+2)^2}{2(h+2)}\right)\\
    \nonumber
   \bk&=(\pi,\pi,\pm k_t^{(-1)}); k_t^{(-1)}=\arccos\left(\frac{\frac{\lambda}{1-\lambda}-1-(h-2)^2}{2(h-2)}\right)\\
     \bk&=(0,\pi,\pm k_t^{(0)}),  (\pi,0,\pm k_t^{(0)}); k_t^{(0)}=\arccos\left(\frac{\frac{\lambda}{1-\lambda}-1-h^2}{2h}\right)
\end{align}
In each of the above, a solution exists only in the region of the $\lambda-h$ plane for which the argument of $\arccos$ lies between $-1$ and $+1$. The regions where these solutions exist are shown by horizontal, vertical and slanted lines respectively in \figref{fig:HopfPD}, main text.

In the following, we focus on the semimetal labelled SM$1$, (\figref{fig:HopfPD}, main text) which separates the $\chi=0$ and $\chi=1$ insulators. The band gap of this semimetal vanishes at $\bk^{\pm}=(0,0,\pm k_t^{(1)})$. A direct computation shows that this semimetal is characterized by a change in the Chern number $C_3$ across the BZ. The corresponding plot for $\lambda=\frac{1}{4}, h=-3$ is shown in \figref{fig:HopfPD} (inset), main text.
\begin{align}\label{eqn:supp_chernchange}
	C_3(k_3) =\left\{
		\begin{array}{cc}
			0 & k_t^{(1)} < \abs{k_3}<\pi \\
			1 & \abs{k_3}<k_t^{(1)} \\
		\end{array}
	\right.
\end{align}
Expanding the Bloch hamiltonian about $\bk^+$, we find that the bands cross linearly: $ H(\bk^+ +\bq) \approx 2 \left(1-\lambda\right)\ba(\bq) \cdot \bsig $ where
\begin{align}
\nonumber
    a_1(\bq)&=q_1 \sin{k_t^{(1)}}+q_2 \left(h+2+\cos{k_t^{(1))}}\right)\\
    \nonumber
    a_2(\bq)&=q_2 \sin{k_t^{(1)}}- q_1 \left(h+2+\cos{k_t^{(1))}}\right)\\
    a_3(\bq)&=q_3 \left(h+2\right) \sin{k_t^{(1)}} 
\end{align}
Note that, for the semimetal corresponding to a point on the boundary between the $\chi=1$ insulator and SM$1$, the coefficient of $q_3$ in the Taylor expansion above vanishes, and the bands touch quadratically (along the $k_3$ direction) at $\bk^{\pm}=(0,0,0)$ 

\section{Hopf semimetals}\label{sec:supp_hopfsemimetals}
\begin{figure*}
    \includegraphics[width=0.45\textwidth]{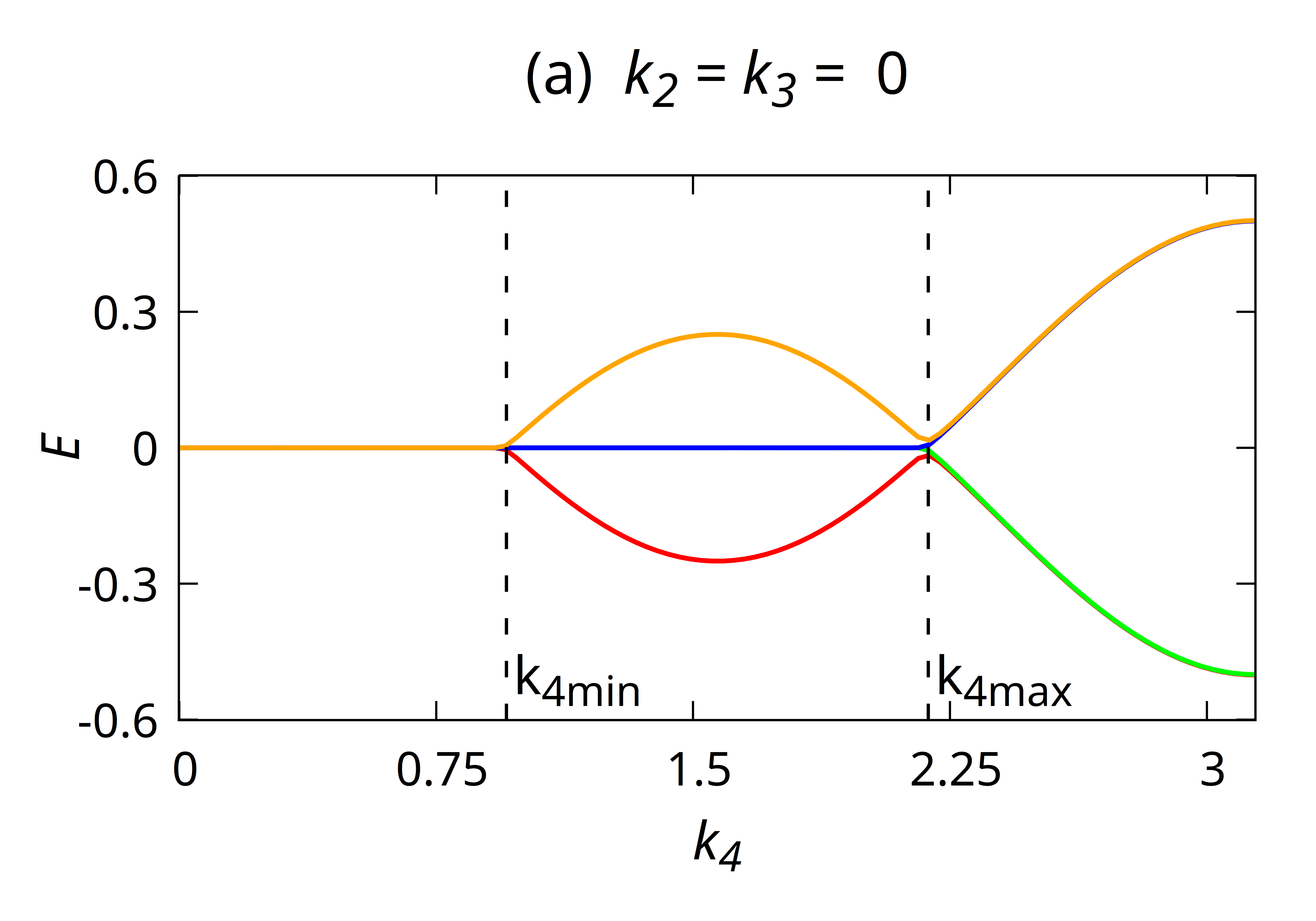}
    \includegraphics[width=0.45\textwidth]{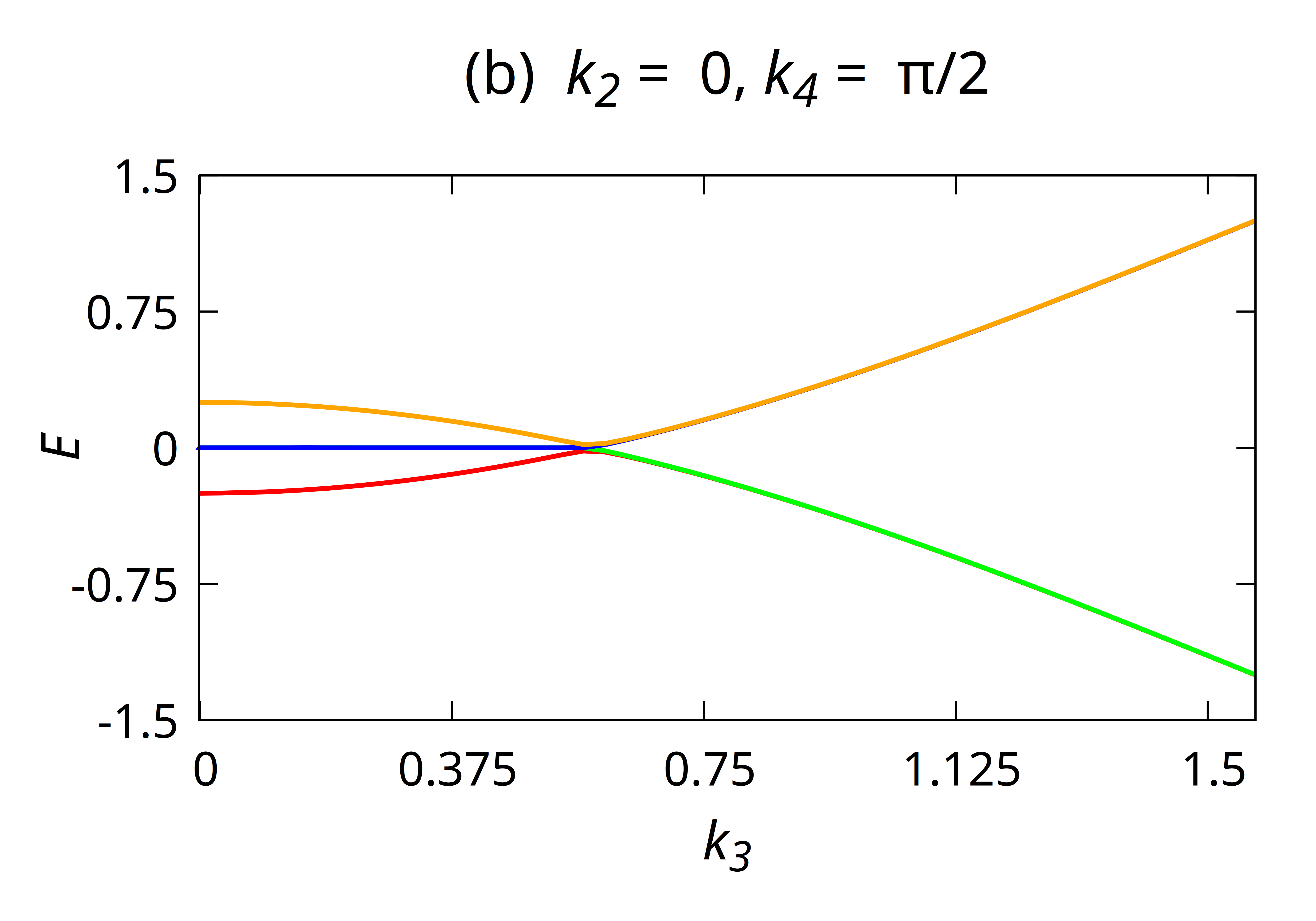}
    \includegraphics[width=0.45\textwidth]{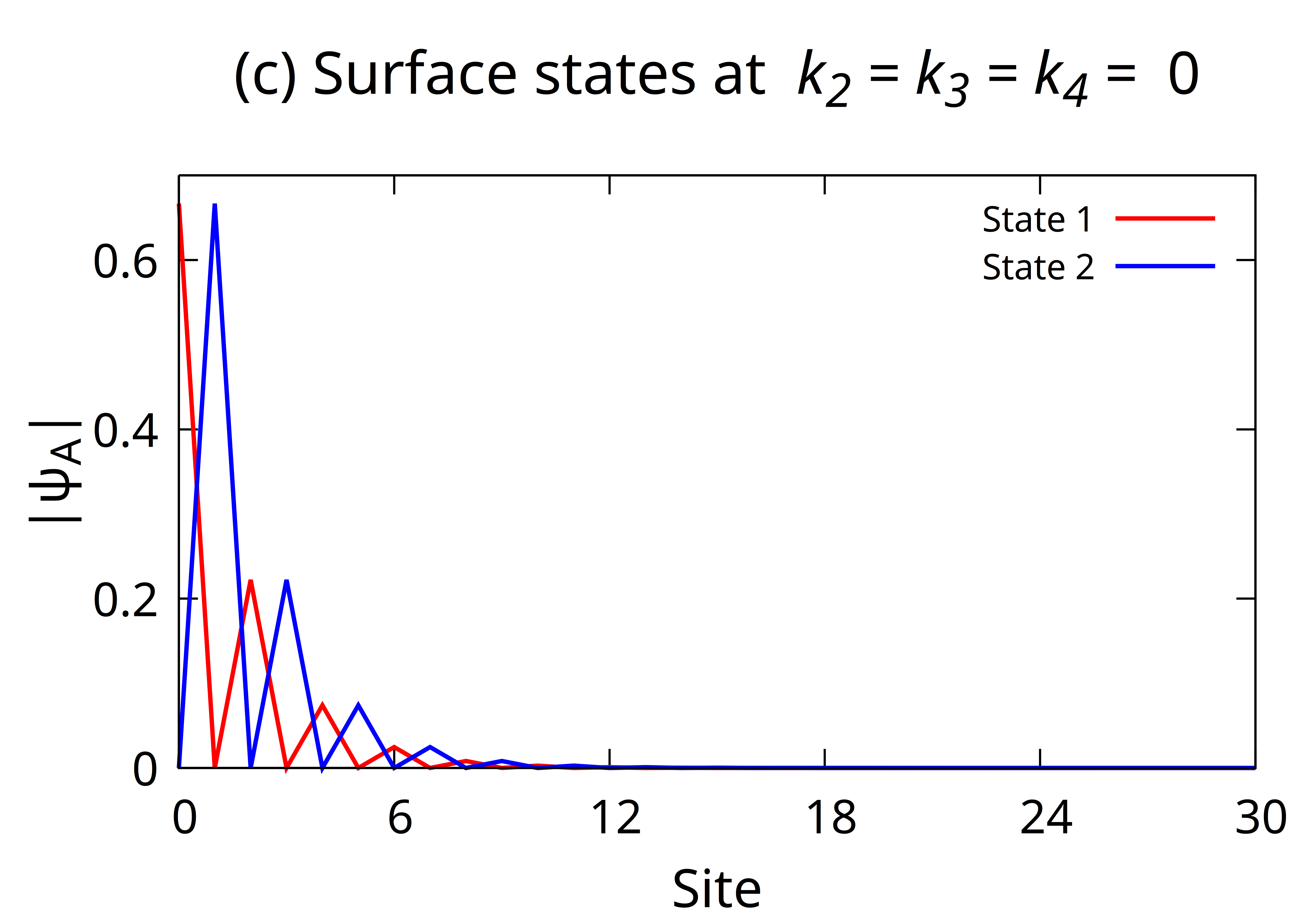}
    \caption{Surface states of the surface normal to $(1,0,0,0)$ on an $L=256-$site thick slab. Energies of the two bands above and the two below the Fermi energy are shown. (a) Band dispersion for $k_2=k_3=0,k_4 \in [0,\pi)$. (b) Band dispersion for $k_2=0, k_4=\frac{\pi}{2}, k_3 \in [0,\frac{\pi}{2}]$. (c) the magnitudes of wavefunctions on $A$ orbital for the two fermi-arc states at $k_2=k_3=k_4=0$, localized at the same end. Even though $L=256$, wavefunctions are shown only on the first thirty sites for clarity. }
    \label{fig:supp_surface1000} 
\end{figure*}
We construct a two-band semimetallic system on a cubic lattice in four dimensions with $0<\lambda<\frac{1}{2}$ as a tuning parameter. Let us denote $\bK=(\bk, k_4) \in T^4$ and $\bk=(k_1,k_2,k_3)$
\beq\label{eqn:supp_HopfNLSM}
\bd(\bK,\lambda) = (1-\lambda) \bd^{(1,1)}(\bk,-3 + \cos{k_4}) + \lambda \bd^{\textup{f}}(\bk) 
\eeq
When $\lambda=0$, we obtain a semimetal with bands touching at exactly two points $\bK^{\pm}=(0,0,0,\pm \frac{\pi}{2})$. Expanding the hamiltonian to lowest order near the band touching points reveals the interesting topological aspects of this semimetal. Writing $H(\bK^{\pm} + \bQ) \approx \ba^{\pm}(\bQ) \cdot \bsig $, where $\bQ=(q_1,q_2,q_3,q_4)$
\begin{align}
a_{1}^{\pm}(\bQ)&=2(q_1 q_3 \mp q_2 q_4)\\
a_{2} ^{\pm}(\bQ)&=2(q_2 q_3 \pm q_1 q_4)\\
a_{3} ^{\pm}(\bQ)&=q_1^2 +q_2^2 -q_3^3 -q_4^2
\end{align}
Note that these are the well-known Hopf maps with Hopf indices $-1, +1$ respectively. This beautifully demonstrates that the points $\bK^{\pm}$ are sources of ``Hopf flux". Consider a ball $B^+$ centered at the point $\bK^+$. Its boundary $\partial B^+$, homeomorphic to $S^3$ encloses the point. Now, the restriction of the hamiltonian \ref{eqn:supp_HopfNLSM} to $\partial B^+$ defines a map from $S^3$ to $S^2$. From the above, we see that it belongs to the homotopic class labelled by the Hopf index $\cH=-1$. Similarly, the restriction of the hamiltonian to a surface homeomorphic to $S^3$, and one which encloses the point $\bK^-$ defines a map with Hopf index $\cH=+1$

As discussed in the main text, point touching in a two-band system in four dimensions is unstable, and the band touching points evolve to nodal lines $L^{\pm}$ for an infinitesimal $\lambda>0$. Importantly, any surface homeomorphic to $S^3$ and enclosing the nodal lines $L^{\pm}$ still carries the Hopf index $\cH=\mp 1$. It is easily seen that the nodal lines lie in the $k_3-k_4$ plane (i.e., $k_1=k_2=0$) of the BZ. The equation of the nodal lines can be obtained by the replacement $k_t^{(1)} \rightarrow k_3$, $h \rightarrow -3 +\cos{k_4}$ in \eqnref{eqn:supp_kt_solns}. We find that nodal lines are given by the curves
\begin{equation}\label{eqn:supp_nodalline_eqn}
2(1-\cos{k_3})(1-\cos{k_4})+\cos^2{k_4}=\frac{\lambda}{1-\lambda}
\end{equation}
The nodal line lying in the region $k_4>0$ extends from $k_4^{\textup{min}} \leq k_4 \leq k_4^{\textup{max}}$. Observing that $k_3=0$ when $k_4=k_4^{\textup{min}}$ or $k_4=k_4^{\textup{max}}$, we obtain
\begin{align}\label{eqn:supp_k4_minmax}
k_4^{\textup{min}}&=\arccos{\sqrt{\frac{\lambda}{1-\lambda}}}\\
k_4^{\textup{max}}&=\pi-\arccos{\sqrt{\frac{\lambda}{1-\lambda}}}
\end{align}
Finally, for the semimetal $\lambda>0$, a Taylor expansion of the Bloch hamiltonian about a point on the nodal line shows that the bands cross linearly expect at the points given by $k_3=0$. Let $\bK^0=(0,0,k_3^0,k_4^0)$ be a point on the nodal line. i.e., $(k_3^0,k_4^0)$ lies on the curve \eqnref{eqn:supp_nodalline_eqn}. Now, taylor-expanding the hamiltonian about this point gives $H(\bK^0+\bQ) \approx 2 (1-\lambda) \ba^0(\bQ) \cdot \bsig $ with $\bQ=(q_1,q_2,q_3,q_4)$ and
\begin{align}
    \nonumber
    a_1^0(\bQ)&=q_1 \sin{k_3^0}+ q_2 \left(\cos{k_3^0}+\cos{k_4^0}-1\right)\\
    \nonumber
    a_2^0(\bQ)&= q_2 \sin{k_3^0}- q_1 \left(\cos{k_3^0}+\cos{k_4^0}-1\right)\\
    a_3^0(\bQ)&=q_3 \sin{k_3^0} \left(\cos{k_4^0}-1\right) +q_4 \sin{k_4^0} \left(\cos{k_3^0}+\cos{k_4^0}-1\right)
\end{align}
At a given point on the nodal line, $\ba(\bQ)$ must vanish along the tangent to the curve in the $k_3-k_4$ plane, and the same is seen from above.
\section{Surface states in Hopf semimetals}
\label{sec:supp_surfacestates}
We have seen that the $T^3$ submanifolds of the BZ of the Hopf semimetal carry a Hopf invariant $\chi=1$ for $\abs{k_4}<k_{4 \textup{min}}$. Hence, we expect Fermi-arc like surface states on the surfaces normal to $(1,0,0,0)$, $(0,1,0,0)$ and $(0,0,1,0)$, from the fact that the three-dimensional Hopf insulators themselves host surface states. To investigate the nature of surface states on the surface normal to $(1,0,0,0)$ direction, we consider a system of finite size with open boundary conditions (slab) in the $(1,0,0,0)$ direction. This explicitly breaks the lattice translation symmetry along the $(1,0,0,0)$ direction, but crystal momenta $k_2, k_3,k_4$ are still good quantum numbers. Thus, the Brillouin zone of the surface is $T^3$. The real space hoppings along $(1,0,0,0)$ can be found by inverse-Fourier transforming the Bloch hamiltonian, and the spectrum of single-particle energies is thus obtained. Assuming that the slab is $L-$site thick, with two orbitals $A,B$ per site, we have $2L$ bands in the BZ $T^3$.

The $(1,0,0,0)$ surface has ``Fermi-arc" surface states for $k_2=0, k_3=0, \abs{k_4}<k_{4 \textup{min}}$. On this fermi-arc, there are four surface states (two orthogonal states localized on each end of the slab), which correspond to the $(1,0,0)$ surface states of the three-dimensional Hopf insulator (\ref{eqn:supp_T3toS2}, $p=q=1$). The wavefunctions of fermi-arc states at the point $k_2=k_3=k_4=0$ are shown in \figref{fig:supp_surface1000}(c). In addition to these, there are two surface states at every point interior to the curve bounded by the projected nodal lines in the $k_3-k_4$ plane. They correspond to the edge states of the $1-2$ submanifolds of $T^4$ which have non-zero Chern numbers for $k_{4 \textup{min}}<\abs{k_4}<k_{4 \textup{max}}$. The energies of the bands just above and below the Fermi energy as $k_4$ is varied, at $k_2=k_3=0$, are shown in \figref{fig:supp_surface1000}(a). In \figref{fig:supp_surface1000}(b), the band dispersion is shown as $k_3$ is varied at $k_2=0,k_4=\frac{\pi}{2}$.
\begin{figure*}
\includegraphics[width=0.45\textwidth]{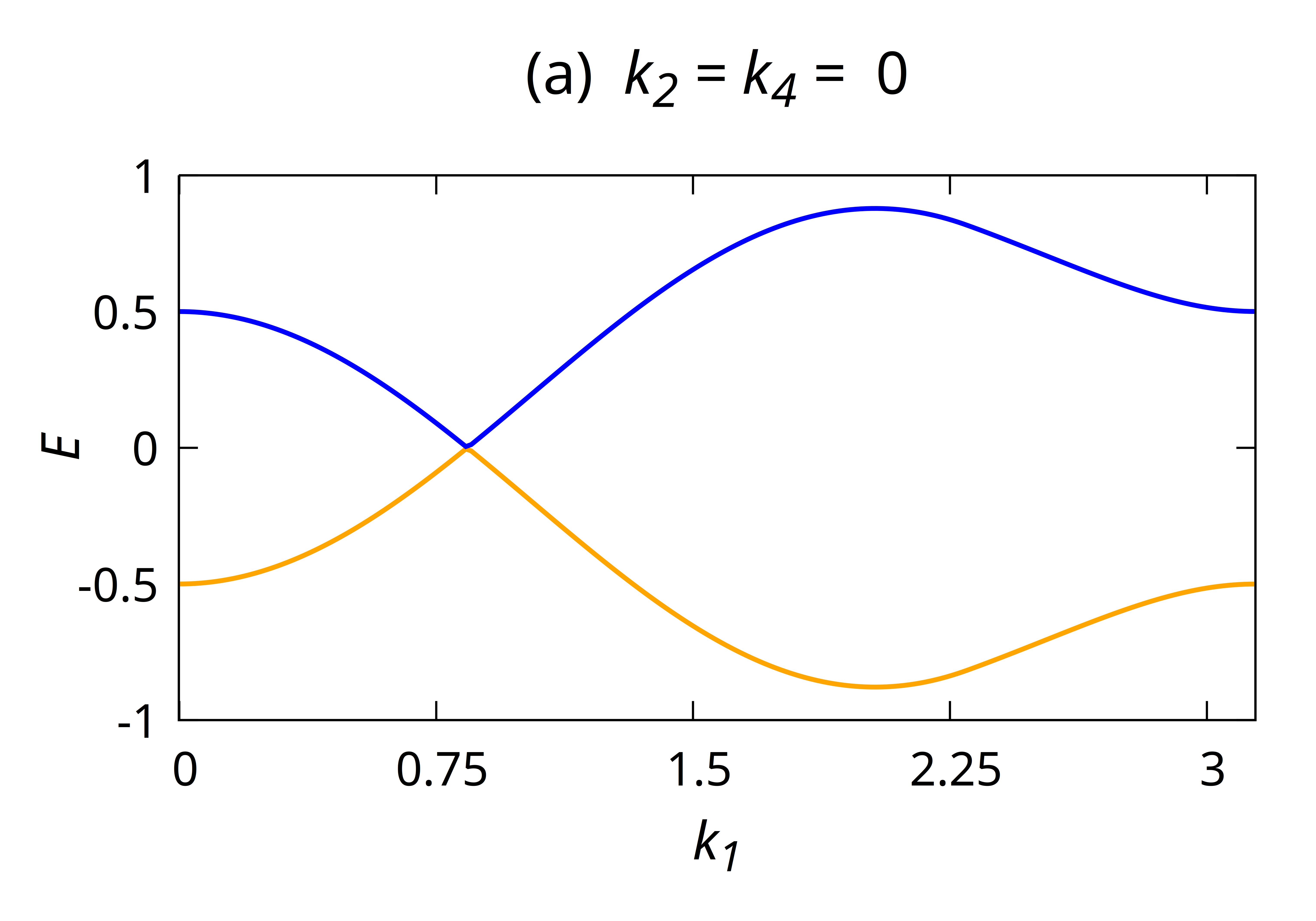}
\includegraphics[width=0.45\textwidth]{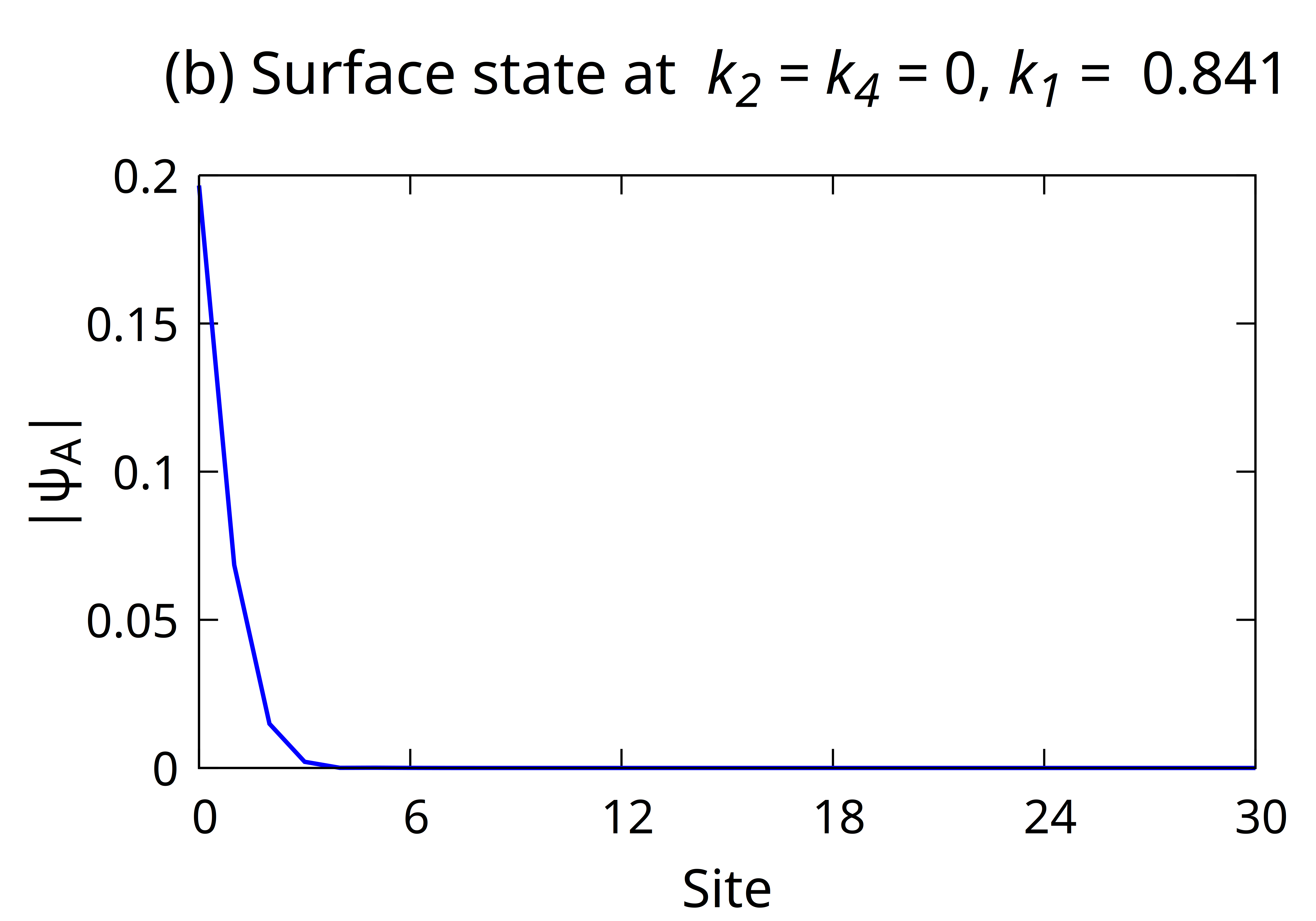}
\caption{Surface states of the surface normal to $(0,0,1,0)$ surface on a $L=256-$site thick slab. (a) Dispersion of the band just above and the band just below the Fermi energy  for $k_2=0,k_4=0,k_1 \in [0,\pi)$. (b) The magnitude of the surface state wavefunction on $A$ orbital at the point $(k_1,k_2,k_4)=(0.841,0,0)$ on the Fermi-surface. $L=256$, but the wavefunction is shown only on the first thirty sites.}
\label{fig:supp_surface0010}
\end{figure*}
The nature of surface states on the surface normal to $(0,1,0,0)$ is similar to those on the $(1,0,0,0)$ surface discussed above. Analogous computation for the $(0,0,1,0)$ surface reveals that there is ``Fermi-surface" of surface states which joins with the nodal lines projected on to the surface BZ, as shown in \figref{fig:Hopf_surface_dir3}, main text. There are two surface states on each point of the fermi-surface, with one state localized on each end of the slab. The band energies for $k_2=k_4=0$, as $k_1$ is varied, are shown in \figref{fig:supp_surface0010}(a). The surface state wavefunction for $(k_1,k_2,k_4)=(0.841,0,0)$ (a point on the Fermi-surface) is also plotted in \figref{fig:supp_surface0010}(b).

\fi

\end{document}